# Thermal cycling effects on static and dynamic properties of a phase separated manganite

J. Sacanell [1,2,*], B. Sievers [1], M. Quintero [1,2,3], L. Granja [1,2], L. Ghivelder [4] and F. Parisi [1,3]

[1] Departamento de Física de la Materia Condensada, GIyA, GAIANN, CNEA, Av. Gral. Paz 1499, San Martin (1650),Buenos Aires, Argentina

[2] Instituto de Nanociencia y Nanotecnología, CNEA-CONICET, Av. Gral. Paz 1499, San Martín (1650), Buenos Aires, Argentina

[3] Escuela de Ciencia y Tecnología, Universidad Nacional de General San Martín, Alem 3901, San Martín (1650), Buenos Aires, Argentina

[4] Instituto de Física, Universidade Federal do Rio de Janeiro, Rio de Janeiro, RJ, 21941-972, Brazil

In this work we address the interplay between two phenomena which are signatures of the out-of-equilibrium state in phase separated manganites: irreversibility against thermal cycling and aging/rejuvenation process. The sample investigated is $La_{0.5}Ca_{0.5}MnO_3$, a prototypical manganite exhibiting phase separation. Two regimes for isothermal relaxation were observed according to the temperature range: for T > 100 K, aging/rejuvenation effects are observed, while for T < 100 K an irreversible aging was found. Our results show that thermal cycles act as a tool to unveil the dynamical behavior of the phase separated state in manganites, revealing the close interplay between static and dynamic properties of phase separated manganites.

*Corresponding author: Joaquín Sacanell
Departamento de Física de la Materia Condensada, Gerencia de Investigación y Aplicaciones, Centro Atómico Constituyentes, Comisión Nacional de Energía Atómica.
Avenida General Paz 1499, San Martín (1650) Buenos Aires, Argentina.
+541167727657
sacanell@tandar.cnea.gov.ar

# I - Introduction

Phase separated (PS) manganites [1] have been the focus of extensive research during the last 15 years due to the strong interplay between their electric, magnetic and structural degrees of freedom. This gives rise to a wide variety of physical properties with many potential technological applications, such as the colossal magnetoresistance effect (CMR) [1] and the nonvolatile memory effect [2,3,4,5], among others. The PS state of manganites is characterized by the coexistence of different magnetic phases and it is the key to understand the behavior of these compounds. Several mechanisms are claimed to be responsible for the appearance of phase separation in manganites. The existence of a Griffiths-like phase [6,7] or the formation of FM polarons [8,9] facilitated by the Jahn-Teller effect of $Mn^{3+}$ ions, are thought to explain inhomogeneities near paramagnetic (PM) to ferromagnetic (FM) transitions. Also, the presence of different kinds of disorder can give rise to the coexistence of longer scale phases. For example, in $(La_{1-y}Pr_y)_{1-x}Ca_xMnO_3$ [10] disorder is induced by cationic substitution of La by Pr and in $La_{0.5}Ca_{0.5}MnO_3$ (LCMO), long range ordering can be suppressed by decreasing grain size [11].

Due to the presence of disorder, manganites with phase separation are typically out of equilibrium in laboratory conditions. For that reason, an evolution of their physical properties evidenced by changes in the relative fraction of the coexisting phases is observed within the PS temperature range [2]. Non equilibrium usually manifest in two main forms. On one hand, an isothermal evolution of the relative fraction of the coexisting phases can be observed, resembling the rejuvenation process of glassy systems [2,12,13,14,15]. In fact, a direct image of the evolution of the coexisting phases has been recently reported using microwave impedance microscopy [16], and confirming the phase diagrams previously obtained by macroscopic measurements [12,17,18], showing that the particular state of PS manganites is strongly dependent on the thermal and magnetic history [12,17,19,20] On the other hand, different systems evolve towards equilibrium by irreversible changes that arise each time the sample is cooled across a transition temperature [21,22,23,24], in close similarity with the phenomenology of the martensitic transitions [25] and shape memory alloys [26]. In this thermal cycling effect, the system evolves from cycle to cycle by a cumulative growth of one phase against the other. The effect is probably related with a release of the disorder produced by the accommodation strain between the coexisting phases that promotes the formation of the most stable low temperature phase.[27]

A close interplay between static and dynamic properties is evident in the above described scenario, and therefore the assessment of both aspects is needed in order to fully characterize the system properties. In this work we perform the study of static and dynamic properties of the well-known PS manganite $La_{0.5}Ca_{0.5}MnO_3$. The system is paramagnetic (PM) at room temperature, presents a transition to a mainly ferromagnetic (FM) state on cooling below $T_C \sim 225K$ and then to a charge-ordered (CO) state at $T_{co} \sim 150K$, which is also antiferromagnetic (AF). However, the second transition is not completed and a fraction of the FM phase remains trapped in the CO host, giving rise to phase separation. Dynamical behavior of the compound is investigated for temperatures below $T_{co}$ through the isothermal evolution of the magnetization within the PS state. Additionally, we studied the effects of successive thermal cycles between 300 K and 50 K and observed its influence on the static and dynamic properties of the compound. The observed behavior is consistent with the reduction of the structural disorder after each cycle, bringing the system closer to an equilibrium state. This fact partially modified the dynamic behavior of the system, giving rise to a new insight regarding the close relation between static and dynamical behavior. This interplay is reflected in the relation between blocking (freezing) temperature and low temperature magnetization.

**II - Experimental**

Polycrystalline samples of $La_{0.5}Ca_{0.5}MnO_3$ (LCMO) were synthesized by the liquid mix technique, using 99.9% purity reactants (oxides and soluble salts). We followed the route detailed in reference [11] to obtain a sample with an average grain size of 450 nm. Chemical composition and crystalline structure were verified by EDS and X ray diffraction, respectively. Magnetization measurements were performed in a commercial vibrating sample magnetometer Versalab™ manufactured by Quantum Design.

**III - Results and Discussion**

Figure 1a displays the magnetization of the LCMO sample. The measurements were carried out between 300K and 50K at 2 K/min, with an applied magnetic field (*H*) of 0.1 T. Below room temperature a PM behavior is observed, followed by a PM to FM transition at $T_C \sim 225K$. On further cooling a reduction of the magnetization is observed at temperatures below 150K, indicating the appearance of the CO-AF phase. The value of the magnetization (*M*) at low temperatures is intermediate between those expected for a fully FM sample and a fully AF sample, evidencing the low temperature PS state. The cooling and warming curves merge from the lowest temperature to around 70 K and on further warming a large thermal hysteresis is observed.

The dynamical behavior of the sample is summarized in Fig. 1b, where we plot the magnetic viscosity (*S*) obtained from the time dependence of the magnetization, *M* vs. *t* . The measurements were performed after field cooled cooling to the target temperature, and then leaving the system to relax for two hours. The isothermal and isofield magnetization measurements as a function of time at different temperatures were analyzed considering that *M* evolves with time following the expression $M(T,t) = S \ln(t/t_0) + M_0(T)$ $M(T,t) = S \ln(t/t_0 + 1) + M_0(T)$ [12] where $t_0$ is a characteristic time scale of the relaxation and $M_0$ is the initial magnetization. The *S* vs. *T* curve observed in Fig. 1b displays two distinctive peaks at around 130 K and at 80 K, indicating the existence of two different dynamical processes within the PS regime. One is associated with the FM to AFM transformation occurring mainly between T=140 and 120 K. This relaxation process diminishes as the temperature is lowered, and eventually disappears around 100 K, either because the system approaches equilibrium, or because of the thermally activated nature of the relaxation process, or both. The reentrant relaxation behavior appearing below 100 K is also related to the transformation from a FM to a non-FM phase, but it should probably correspond to a different relaxation process than the mechanism occurring at higher temperatures. Thus, at least one of the low temperature phases has different characteristics than those coexisting at higher temperatures (T > 100K). This scenario implies the existence of at least three phases involved in the PS system, as already pointed out and sustained by other authors. [27,28,29]

In order to gain some additional insight on the static and dynamic processes involved, different experiments were performed. In the first one we explore the dynamical behavior of the sample after successive thermal cycles between 300 K and 50 K. It is known that phase separated LCMO presents instability against thermal cycling [22,23,24], which is evidenced by the decrease of the low temperature magnetization as a function of the cycle number. In Fig. 2a we sketch the *M* vs. *T* data obtained for different cycles, in which a monotonous reduction of the magnetization of the sample is observed as a function of the cycle number *n* in a wide range of temperatures. In Fig. 2b we present the results of the magnetic viscosity *S* as a function of temperature. It is the same type of data as shown in Fig. 1b, but now after subjecting the sample to several tens thermal cycles. The outcome is that the relaxation peak at T=130K remains almost unaltered, while the lower temperature peak (T ≈ 80K) is now nearly vanished. This lack of low temperature relaxation after thermal cycling shows that the sample is able to reconfigure its low temperature state, getting closer to equilibrium state after each cycle is performed. This is accompanied by a reconfiguration on its structural and magnetic properties, as it is was shown for example by magnetostriction measurements in phase separated manganites [30]. This scenario is compatible with an accumulative accommodation strain occurring each time the low temperature state is reached.

In order to visualize the interplay between static and dynamic properties, we present an analysis obtained from the data of Fig. 2a. Figure 3 displays how the magnetization at 50 K changes as a function of the number of cycles, *n*. The results show that $M_{50K}$ decays almost exponentially within the range of cycles performed. We also present the results of the blocking temperature $T_B$ (marked with open circles in Fig. 2a), defined as the temperature at which cooling and warming magnetization curves separates. This irreversibility point between cooling and warming data signals the limit between a low temperature blocked and the unblocked PS states [12]. The existence of this blocked regime suggests that at least one of the phases alluded previously is a blocked or frozen FM state, regarding that relaxations are towards a decrease of magnetization. A correlation between the lowering of the $M_{50K}$ and the increase of $T_B$ can be seen in the inset of Fig. 3.

The observed phenomena can be described with a simple decay model for the magnetization. At low temperatures, the sample is a mixture of FM and non-FM regions at a nanoscale, within each single crystalline grain [31]. Such coexistence would lead to the appearance of interface effects due to the small structural differences between the coexisting phases, and thus the interface will be subjected to stress. In each cycle the material is *trained* to relax the interface strain and thus at the next cycle the *trained* interface regions will be converted to the equilibrium phase. The change of one phase into another in a given cycle is proportional to the total amount of phase capable to be converted until the system reaches the equilibrium state. The dependence of the FM fraction ($x$) as a function of the number of cycles $n$ will follow the expression $x(n) = A e^{-n/\lambda} + x_{eq}$, with $\lambda$ representing the constant of decay and $x_{eq}$ the equilibrium fraction of the FM phase. From the data of Fig. 3 we obtained a value of $\lambda = 5.2 \pm 0.3$. Through additional experiments we found $\lambda$ almost independent of the cooling/warming rate at least in the range between 2 and 8 K/min. Regarding the blocking temperature $T_B$, its dependence whit $n$ can be nearly fitted as $T_B(n) = T_{B0}\left(1 - e^{-n/\lambda}\right)$ with the same value of $\lambda$ obtained previously, confirming a close relation between the blocking temperature and the distance of the system to equilibrium, $x_{eq} - x(n)$. As $T_B$ can be related with the thermally activated energy barriers separating coexisting phases, the observed relation between $1/T_B$ and $M_{50K}$ (inset of Fig. 3) is consistent with the picture that energy barriers increase as equilibrium is approached, as suggested in previous works. [12,17]

Finally, we investigate the behavior of the system when combining cycling and relaxation (time dependent) experiments. We performed three $M(T)$ cycles between room temperature and 50 K with a cooling rate of 2 K/min on a virgin sample, without previous cycling. After that, we performed a new cycle with an interruption at a fixed dwell temperature of 130 K for 3 hours, and resumed cooling afterwards. Results are shown in Fig. 4a. We repeated the same process, after several cycles, at a new dwell temperature of 90 K, as shown in Fig. 4b. In Fig. 5 we summarize the evolution of $M_{50K}$ for the entire cycling sequences. The value of $M_{50K}$ for cycles in which the cooling was interrupted is indicated with a triangle (relaxation at 130 K) and a star (relaxation at 90 K). Data is presented in a semi logarithmic scale to emphasize the monotonic behavior within the thermal cycling effects at constant rate and the influence of the isothermal relaxations. We can see that the overall trend of $M_{50K}$ is not affected by the isothermal relaxation at 130 K, even though the magnetization shows a significant time variation at 130 K, as seen in the inset of figure 4a. Also, in the subsequent cycles, the previous tendency of $M_{50K}$ as a function of the cycle number is recovered.

A closer look shows that when the cooling process is resumed after stabilization at 130 K, the magnetization curve merges smoothly with the data at low temperatures. In contrast, for the isothermal relaxation at 90 K, a clear step appears between the data at low temperatures and the expected tendency without the intermediate temperature stabilization (see inset of figure 4b). The effect of the isothermal relaxation at 130 K is reminiscent to the rejuvenation effect that is found in glassy systems and it has also been observed in PS manganites [2]. On the other hand, the isothermal relaxation at 90 K has a different character, as a marked reduction of $M_{50K}$ is observed in comparison with the one expected for a non-stop thermal cycle process. In addition, the immediately subsequent cycle does not recover the initial tendency, presenting a variation that is equivalent to some additional intermediate thermal cycles; after that, the previous slope in the $M_{50K}$ vs. *log(n)* graph is recovered (last 3 points in Fig. 5). Thus it is clear that the thermal cycling reflects a process occurring mainly in the low temperature range (< 100K). The evolution of the system is thus clearly influenced by the states explored by the sample, either by thermal cycling or by relaxation. All these facts are consistent with the relaxing strain picture, within a framework dominated by accommodation strain features.

## IV - Conclusions

In summary, we have studied a prototypical phase separated manganite, $La_{0.5}Ca_{0.5}MnO_3$, and showed the close relation between dynamic and static equilibrium properties. Thermal cycling effects, relaxation, aging and rejuvenation experiments reflect how a physical characterization of a PS system must be performed taking into account a whole range of variables, static and dynamic, in the temperature range characterized by phase coexistence. Assisted by thermal cycling effects, the system evolves from one cycle to another through the interface of the FM and CO phases, through a state which retain its memory in the subsequent thermal cycle. Within this picture, the system can be trained by the accommodation strain between the FM and CO phases towards the most stable low temperature state, by a progressive release of structural strain. The role of strain as pinning centers for the interface between phases in coexistence has already been signaled a key point in epitaxially grown thin films [32].

As the system evolves in its nanoscale structural configuration towards an equilibrium state by thermal cycling, the dynamical behavior is affected. The reduction on the magnetic viscosity at low temperatures after several cycles is compatible with the increase of the blocking temperature $T_B$ as accommodation strain is released. Regarding this point, we could identify two distinctive regimes for the dynamical evolution of the physical properties in an LCMO compound. In the higher temperature regime (T around 130 K) the system behavior is characterized by the existence of aging and rejuvenation processes, reminiscent to those occurring in spin glasses. In this regime the system is able to evolve to a closer equilibrium states, but this evolution does not affect the further behavior of the system once the cooling cycle is resumed. This regime can be characterized as dynamic. On the other hand, in the lower temperature regime (below 100K), relaxation allows the system to reach states not accessible through conventional cooling cycles. Furthermore, these new explored states affect the microstructural properties of the sample, and influence the behavior of the system in further cycles. Within this frozen phase separated regime, relaxation occurring below 100 K gradually disappears while increasing the number of cycles, supporting the close relation between structural and dynamic behavior.

The relationship between static and dynamic properties presents several novel aspects which can help to understand the interplay between structural as well as magnetic equilibrium properties, and dynamic behavior in PS manganites. As a main result, the thermal cycling effect results in the release of accommodation strain which in turn affects the behavior of the system on subsequent experiments. Thus, a comprehensive description of any PS manganite must necessarily include static and dynamic characterization of the physical properties under investigation.

**Acknowledgments**

This research was supported by the bilateral cooperation FAPERJ-CONICET, ANPCyT (PICT 1506/2012, 2116/2014) and CONICET (PIP 00362). We thank Pablo Levy and Andrés Arazi for fruitful discussions.

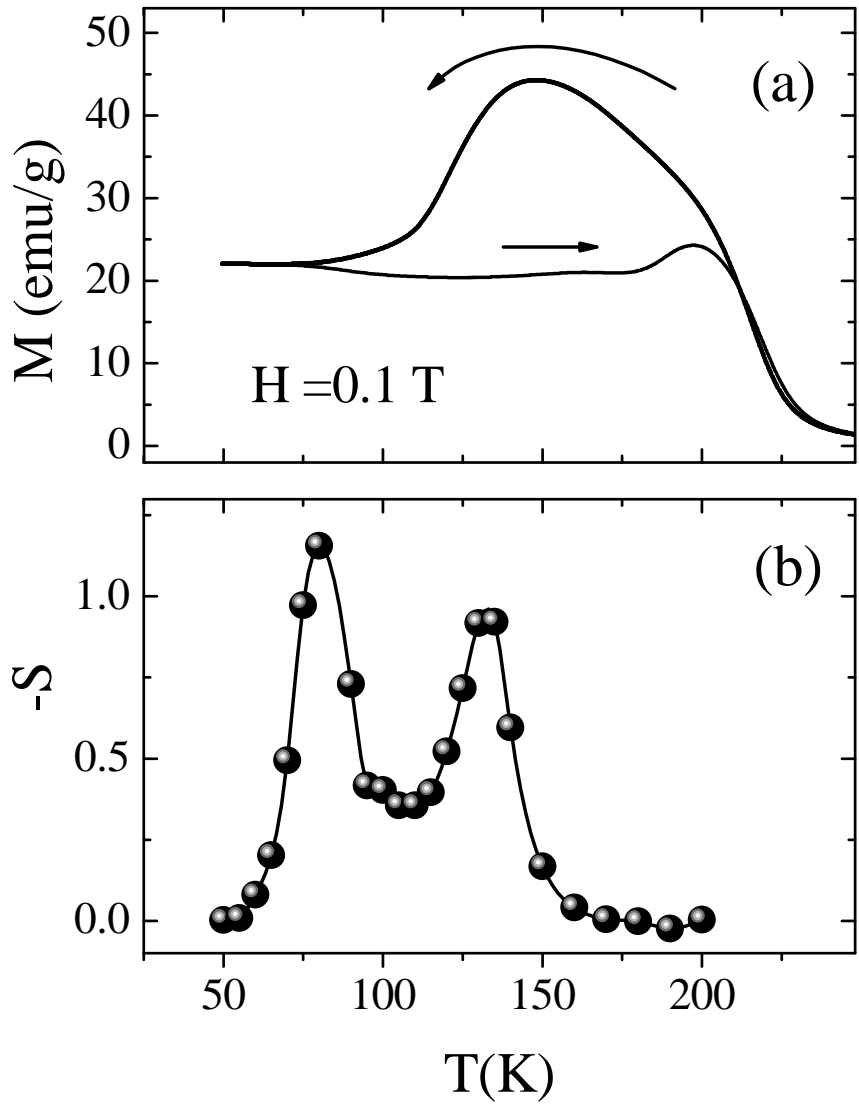

Figure 1: (a) Magnetization vs. temperature of the as prepared $La_{0.5}Ca_{0.5}MnO_3$ sample, measured with an H = 0.1 T. (b) Magnetic viscosity (*S*) as a function of temperature, obtained from time dependent magnetization data (see text for details).

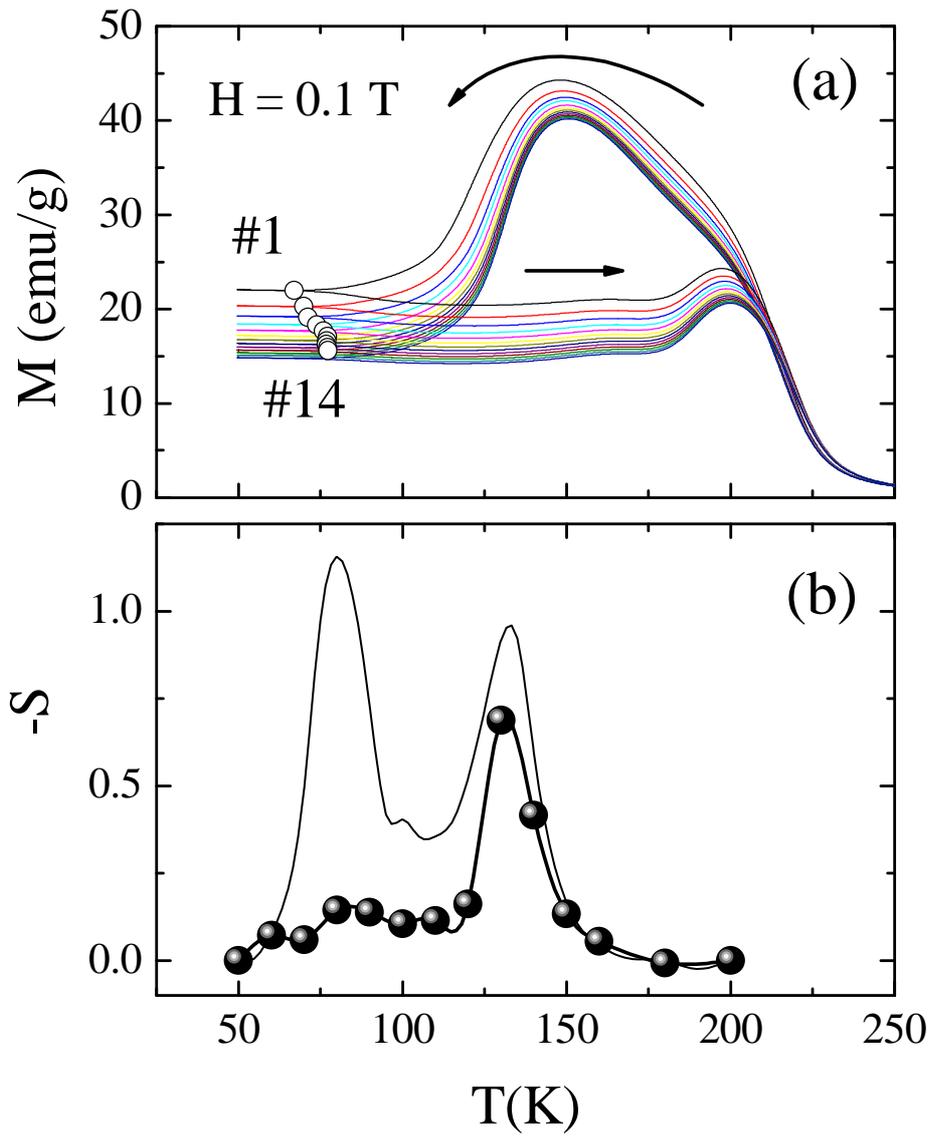

Figure 2: (color online) (a) Magnetization vs. temperature for consecutive thermal cycles. Open circles indicate the low temperature irreversibility point, which define the blocking temperature ($T_B$). (b) Magnetic viscosity ($S$) as a function of temperature after performing several dozens of thermal cycles. The data corresponding to the sample before the sequence of thermal cycles is also presented for comparison.

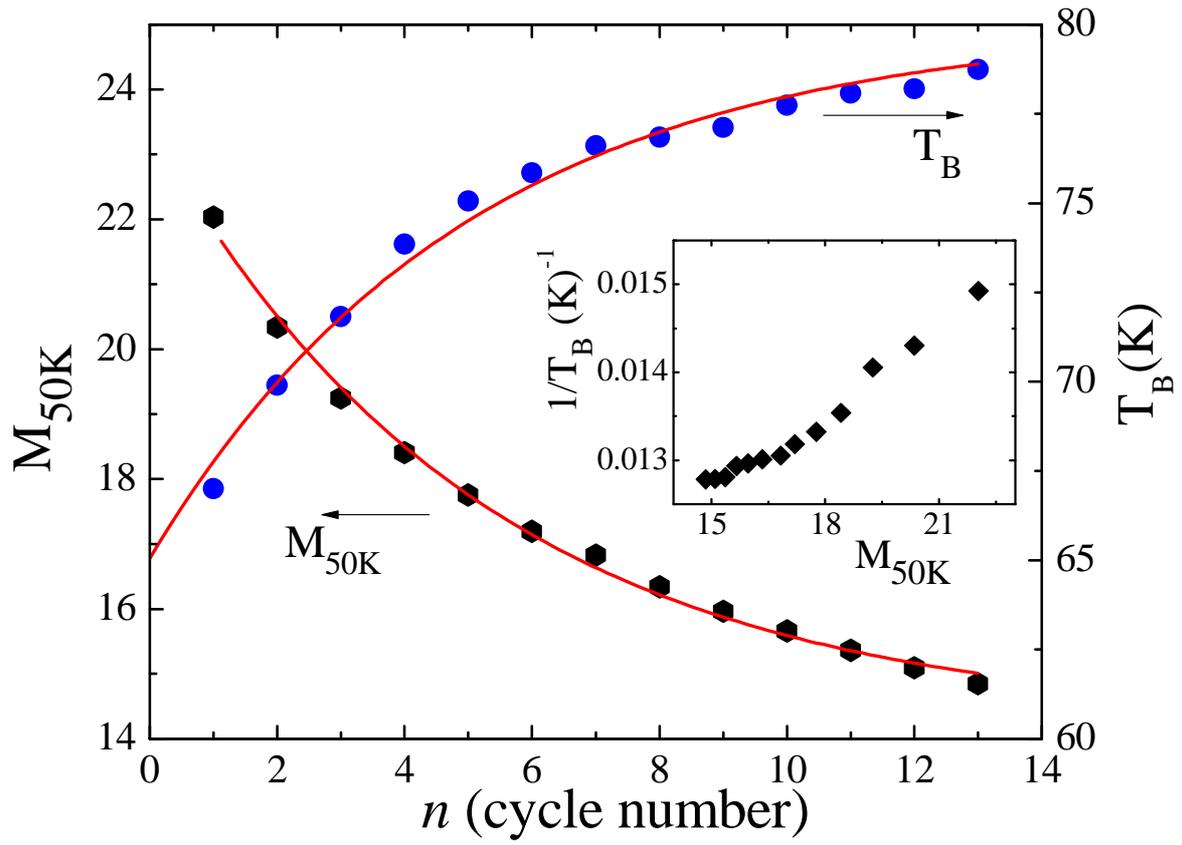

Figure 3: (color online) Magnetization values at 50 K, $M_{50K}$, and blocking temperature, $T_B$, for a sequence of thermal cycles performed to an as prepared sample, as a function of the cycle number $n$. The inset: $1/T_B$ vs. $M_{50K}$.

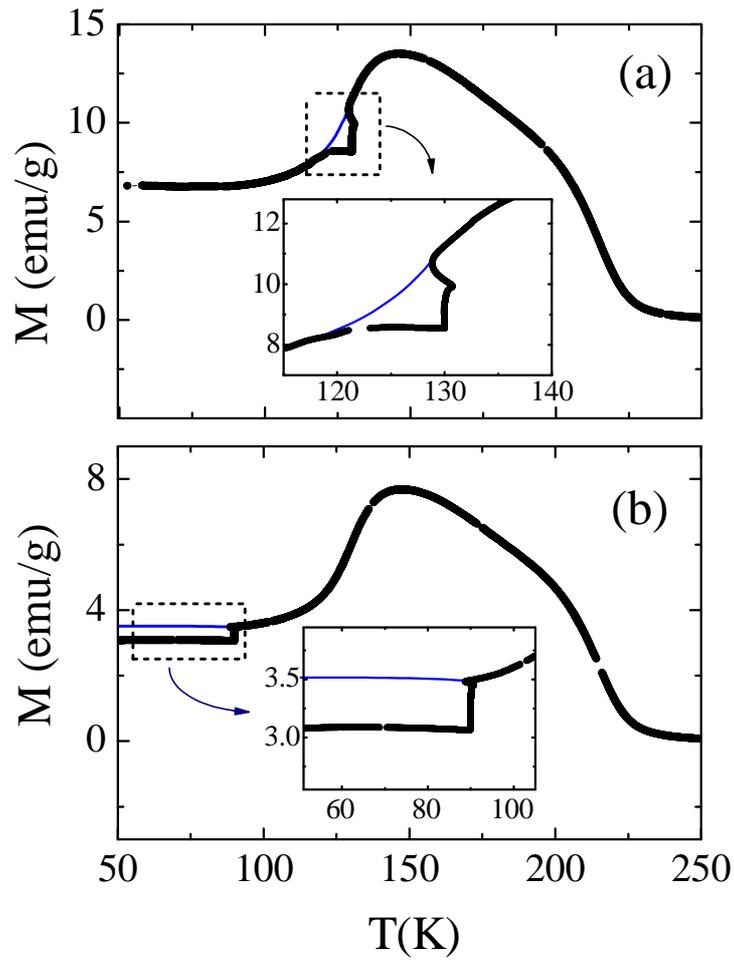

Figure 4: (color online) *M* vs. *T* with isothermal relaxation data at (a) 130 K and (b) 90 K. The blue line indicates the expected tendency without the intermediate time dependent measurements.

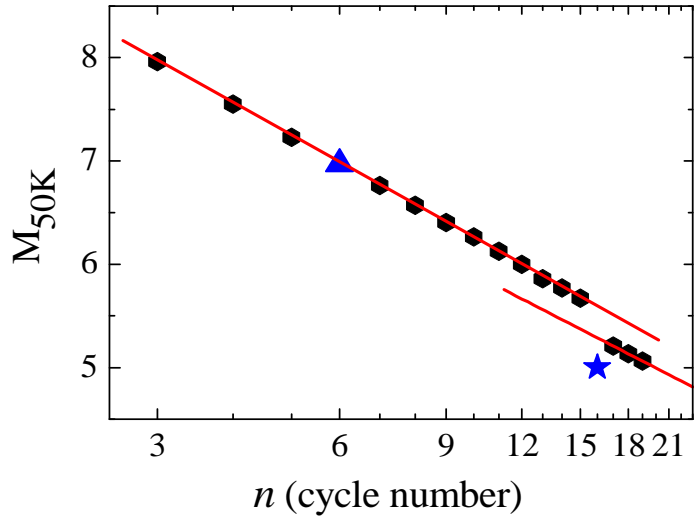

Figure 5: (color online) Magnetization at 50 K as a function of the number of thermal cycles performed, including the two cycles depicted in figures 4a and 4b. $M_{50K}$ for those cycles is marked with a triangle (interruption at 130 K) and a star (interruption at 90 K).